%% file: AAA_NGC4278.tex
\documentclass[]{aastex631}
\usepackage{subfigure}
\usepackage{amsmath}
\graphicspath{{./}{figures/}}

\begin{document}

\title{Discovery of Very-high-energy Gamma-ray Emissions from the Low Luminosity AGN NGC 4278 by LHAASO}

\input{ApJ20240201.tex}

\correspondingauthor{S.C. Hu, G.M. Xiang, M. Zha, L. Chen, T. Wen}

\begin{abstract}

The first source catalog of Large High Altitude Air Shower Observatory reported the detection of a very-high-energy gamma ray source, 1LHAASO J1219+2915. In this paper a further detailed study of the spectral and temporal behavior of this point-like source have been carried. The best-fit position of the TeV source ($\rm{RA}=185.05^{\circ}\pm0.04^{\circ}$, $\rm{Dec}=29.25^{\circ}\pm0.03^{\circ}$) is compatible with NGC 4278 within $\sim0.03$ degree. Variation analysis shows an indication of the variability at a few months level in the TeV band, which is consistent with low frequency observations. Based on these observations, we report the detection of TeV $\gamma$-ray emissions from this low-luminosity AGN NGC 4278. The observations by LHAASO-WCDA during active period has a significance level of 8.8\,$\sigma$ with best-fit photon spectral index $\varGamma=2.56\pm0.14$ and a flux $f_{1-10\,\rm{TeV}}=(7.0\pm1.1_{\rm{sta}}\pm0.35_{\rm{syst}})\times10^{-13}\,\rm{photons\,cm^{-2}\,s^{-1}}$, or approximately $5\%$ of the Crab Nebula. The discovery of VHE from NGC 4278 indicates that the compact, weak radio jet can efficiently accelerate particles and emit TeV photons.

\end{abstract}



\section{Introduction}

Active Galactic Nuclei (AGNs) exhibit diverse emissions across the electromagnetic spectrum, with gamma-ray emissions playing a crucial role in understanding ultra high energy cosmic rays and AGN physics. The very high energy (VHE, $\gtrsim0.1$ \,TeV) gamma-ray sky primarily comprises radio-loud AGNs (see TeVCat, http://tevcat.uchicago.edu/.), particularly blazars, with relativistic jets directed toward us, enhancing emissions due to beaming effects \citep{1995PASP..107..803U}. Of the 88 known VHE AGNs, 82 are blazars (9 FSRQs and 73 BL Lacs) with relativistic jets close to the line of sight, leading to enhanced broadband emissions from relativistic beaming. Blazar gamma-ray emissions stem from either leptonic processes, i.e. inverse Compton scattering by energetic electrons, or hadronic processes involving $\pi^{0}$ meson decay or synchrotron radiation from ultra-relativistic protons. In addition to VHE blazars, there are also 4 VHE radio galaxies (RGs) and two VHE AGNs exhibiting both RG and BL Lac properties: IC 310 and PKS 0625-35 \citep[see TeVCat and][]{2022Galax..10...61R}. According to unified schemes of radio-loud AGNs, RGs are the parent population of blazars, with jet directions misaligned with the line of sight, resulting in less beamed emissions \citep{1995PASP..107..803U}. These detected 4+2 VHE radio galaxies all belong to the FRI subclass, characterized by bright jets in the center with edge-darkened radio structures \citep{1974MNRAS.167P..31F}. The origin of VHE emissions from FRI radio galaxies remains unclear but may be attributed to stratified jets \citep{2005A&A...432..401G}, extended jet emission regions \citep{2020Natur.582..356H}, or the vicinity of black holes \citep{2011ApJ...730..123L}. VHE emissions offer a unique view into the extreme physical processes within AGN jets, particularly in misaligned RGs. Current observations of VHE emissions from AGNs primarily operate in pointing mode, targeting objects during flaring states. The lack of uniform coverage across the VHE sky impacts the completeness of the VHE AGN catalog, especially for low luminosity AGNs (LLAGNs).

LLAGNs are very common in the nearby universe, and occupy the bulk of the fainter end of AGN luminosity population \citep{1999ApJ...516..672H, 2005A&A...435..521N, 2008ARA&A..46..475H}. They exhibit low bolometric luminosity relative to Eddington luminosity ($L_{\rm{bol}}/L_{\rm{Edd}}\sim10^{-6}\text{--}10^{-4}$), lack a prominent big blue bump and Fe K$\alpha$ line, and show flat or inverted radio spectra \citep{1999ApJ...516..672H, 2008ARA&A..46..475H}, with radio variability typically on month-long timescales \citep{2002A&A...392...53N, 2005ApJ...627..674A, 2008ARA&A..46..475H}. Their radio loudness increases at lower luminosities \citep{2007MNRAS.377.1696M, 2008ARA&A..46..475H}, and they are generally more radio-loud than typical AGNs \citep{2005A&A...435..521N}. These systems exhibit radiatively inefficient accretion flow (RIAF) due to low accretion rates \citep{2008ARA&A..46..475H}. Studying VHE emissions from these sources can offer insights into diverse mechanisms responsible for VHE emission in AGN jets, especially regarding jet-environment interactions, and aid in understanding the origin of the $\gamma$-ray background.

The first LHAASO source catalog reported the detection of the VHE gamma ray source 1LHAASO J1219+2915 \citep{lhaaso-catalog-2024}. Because of its possible association with NGC 4278, in this paper a further detailed study of the spectral and temporal behavior of this point-like source have been performed with better statistics.  NGC 4278, a typical low-luminosity AGN classified as a LINER due to its weak H$\alpha$ line, has a central black hole mass of $\approx3\times10^{8}\,\rm{M_{\odot}}$ at distance $D_{\rm{L}}=16.4$\,Mpc \citep{2001ApJ...546..681T}, with an Eddington ratio of $\approx5\times10^{-6}$ \citep{2014A&A...569A..26H}, indicative of RIAF.

The paper is structured as follows: Section 2 introduces the LHAASO detector and dataset. Section 3 details the multi-TeV gamma-ray detection from LLAGN NGC 4278, including spectra and variability analysis. Section 4 compares with other VHE AGNs. Throughout, a $\Lambda$CDM cosmology with Planck results is adopted, with parameters $\Omega_{\rm{m}}=0.32$, $\Omega_{\Lambda}=0.68$, and $H_{0}=67.4$\,km\,s$^{-1}$\,Mpc$^{-1}$.

\section{The LHAASO Experiment and Data}
The Large High Altitude Air Shower Observatory (LHAASO) is located in Haizi Mountain with altitude of 4410m asl, (29$^\circ$21'27.6"\,N, 100$^\circ$08'19.6"\,E) in Daocheng, Sichuan Province, China. It consists of KiloMeter-square Array (KM2A), Water Cherenkov Detector Array (WCDA), 18 Wide Field-of-view Cherenkov Telescope Array (WFCTA). LHAASO is a multi-scientific-purpose Extensive Air Shower (EAS) array designed to detect the cosmic rays and gamma-ray air showers in a wide energy range, from sub-TeV to beyond 1\,PeV. The WFCTA primarily focuses on cosmic ray physics, while the KM2A and WCDA are mainly dedicated to gamma-ray astronomy. The WCDA is sensitive to gamma-rays from sub-TeV to tens of TeV and KM2A is sensitive from tens of TeV to several PeV. With a large detector area and excellent gamma-ray/background discrimination power, the sensitivity of LHAASO for both VHE and UHE gamma-ray observations are much higher than any other EAS experiments. 


The results presented here are mainly obtained using WCDA data taken between 5th March 2021 and  31st October 2023. The effective live-time is about 891\,days. To keep good reconstruction quality the following cuts are applied for both the experimental data and the simulated samples:

\begin{itemize}
\item[1)] events should have at least 60 fired number of detectors, $N_{\rm{hit}}\ge$ 60;
\item[2)] events should have a zenith angle less than $50^{\circ}$, $\theta\le50^{\circ}$; 
\item[3)] for every shower event, a variable called RMDS  was determined as a measurement of shower core location error based on top 10 of the hottest detectors, by requiring RMDS $\le$ 20\,m, better resolution on shower core position could be achieved;
\item[4)] the Gamma/Proton separation parameter, $\mathcal{P}_{\rm{incness}}$, is required to be less than 1.12, 1.02, 0.90, 0.88, 0.84 and 0.84 for segments with  $N_{\rm{hit}}$ value of [60\text{--}100), [100\text{--}200), [200\text{--}300), [300\text{--}500), [500\text{--}800) and  [800\text{--}2000].
\end{itemize}

After applying these cuts, the number of gamma-like events WCDA recorded is around $8.3\times10^9$ events. LHAASO-KM2A data for the same time period are used for this work, with about $1.3\times10^7$ gamma-like events after quality cut. Details about LHAASO-KM2A quality cuts can be found in \cite{2021CPC...45}. Further details about the detector and the reconstruction can be found in \cite{lhaaso-wcda-2021}.

\section{Analysis and Results}

The event and background maps are generated in celestial coordinates (right ascension and declination in epoch J2000.0) with grid size of $0.1^{\circ}\times0.1^{\circ}$. The so called ``direct integration method" \citep{Fleysher:2003nh} is employed to estimate the number of background events.  In this work, the integration time is set to 4 hours, and events within the regions of the Galactic plane ($|b|<10^{\circ}$) and gamma-ray sources (with a spatial size less than 5$^{\circ}$) are excluded from the background estimation. The excess of source map is then obtained by subtracting the background map from the event map.

For WCDA the events are grouped into the six analysis bins based on the effective number of fired PMTs, allowing us to extract the energy spectrum of gamma-ray source. Using Crab Nebula trajectory as a reference, the corresponding energies range from 0.5 to 20\,TeV. To convert the gamma ray count to flux, the detector response is calculated with a simulated data samples. The simulation samples are generated with the air shower simulation code CORSIKA and the detector simulation package G4WCDA \citep{lhaaso-wcda-2021}. The energy of simulated gamma rays is sampled from 1\,GeV to 1\,PeV with Crab Nebula trajectory up to zenith angle of 70$^{\circ}$. For KM2A data, considering the reconstructed energy resolution and statistics, one decade of reconstruction energy, $E_{\rm{rec}}$, is divided into 5 bins with a width of $\textrm{log}_{10}E_{\rm{rec}}=0.2$ for event and background map.

A multi-dimensional maximum likelihood analysis based on forward folding method \citep{2021CPC...45} is applied to fit the excess maps to estimate position and photon flux of the sources with 5 free parameters: right ascension $\alpha$, declination $\delta$, extension $\sigma$, differential flux $\phi_{0}$ at $E_{\rm{piv}}$ and photon spectral index $\varGamma$. While propagating towards the Earth, extragalactic VHE photons may be absorbed by the extragalactic background light (EBL) through the pair production process ($\gamma\gamma\rightarrow e^{+}e^{-}$). In this work a power-law spectrum with attenuation ${\rm d}N/{\rm d}E=\phi_{0}(E/E_{\rm{piv}})^{-\varGamma}e^{-\tau(E)}$ is assumed as the sources observed spectral energy distribution. The reference energy $E_{\rm{piv}}$ is fixed at 3~TeV, $\tau(E)$ is the photon attenuation derived from the EBL model \citep{2021MNRAS.507.5144S} as a function of energy at the distance $D_{\rm{L}}=16.4$\,Mpc \citep{2001ApJ...546..681T}.
A likelihood ratio test is performed on a test statistic,
$\rm{TS} = -2(\ln \mathcal{L}_{\rm{b}} - \ln \mathcal{L}_{\rm{s+b}})$. 
Here $\ln\mathcal{L}_{\rm{s+b}}$ is the signal plus background model, $\ln\mathcal{L}_{\rm{b}}$ is background-only model. Then TS is numerically maximized by iteratively varying the input parameters. It should be pointed out that for the light curve estimation, the spectral index was kept constant and only the differential flux value was left free to vary in the likelihood maximization.

\subsection{Light Curve}

\begin{figure}
\centering
\includegraphics[width=0.85\textwidth]{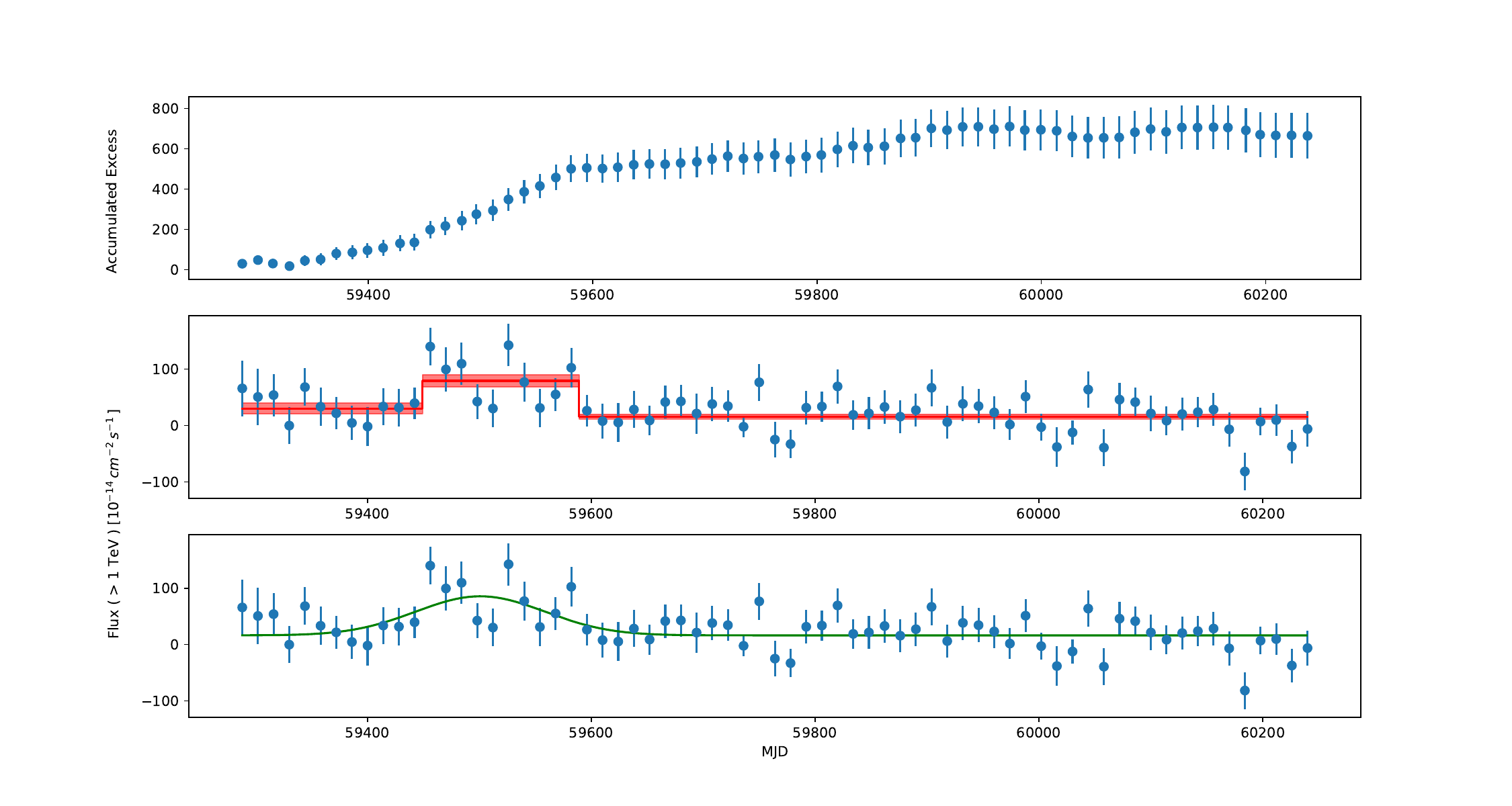}
\caption{The light curves of NGC 4278. Top panel is the cumulative excess light curve, the middle and bottom panels are flux light curve. Red line in middle panel is the result from bayesian blocks analysis, green line in bottom panel is the fitting result from a combined Gaussian function, the parameters are reported in the text.}
\label{figngc4278_cl2}
\end{figure}

Top panel of Figure \ref{figngc4278_cl2} shows the accumulated excess as a function of time from which one can find that a rapid change in the slope is obvious, indicating the existence of a variable phenomenon. In order to further quantitatively estimate this variable behaviour, 
we have binned the light curve into two-week interval (14 transits), as shown in middle panel of Figure \ref{figngc4278_cl2}. The Bayesian blocks algorithm \citep{bayesian-block-2013} is applied to identify the optimal source states. An active state, which started on MJD 59449 and ended on MJD 59589 with a duration of 140 days, has been identified. It is depicted as a red block curve in the top panel of the same plot. It should be noted that negative flux points are shown in this figure, even though they are not physical. This occurs when low statistics lead to an under-fluctuation of the event count relative to the background estimation, which also explain the occasional decline in the cumulative count. The significance of the dip around MJD = 60184 is $-2.6\,\sigma$, which is equivalent to a post-trial probability of 0.55 ($-0.58\,\sigma$). 

Additionally, based on the likelihood variability test \citep{hawc-daily-2017}, we conducted the calculation to determine the probability of a given source having a constant flux, resulting in a TS value of 105.1, which is corresponding to a p-value of $2.6\times10^{-3}$ and indicates a variable nature of the TeV emission from this source. Notably, the pre-trial significance map within the variable duration has reached 8.8 standard deviations (as shown in the next section). This is significantly higher than a stable source behavior during the same time period, which is less than 2 standard deviations.

In addition, we also adopted another way to fit the bi-weekly light curve. The green line in the bottom panel of the figure displays the fitting result using a combination of a constant term and a Gaussian function to describe the variable behavior in time, i.e. $F(t)=F_{\rm{c}}+F_0\times e^{-(t-t_0)^2/2\tau_{\rm{s}}^2}$, the best fitting parameters are $t_0$=59501~$\pm~15$, $\tau_{\rm{s}}=58~\pm~16$\,days, $F_{\rm{c}}=(16.3~\pm~4.5)\times10^{-14}\,\rm{cm^{-2}s^{-1}}$, denoting the stable amplitude and $F_0=(69.4~\pm~16.6)\times10^{-14}\,\rm{cm^{-2}s^{-1}}$, denoting the variable amplitude. 
It is interesting to note that the time scale width obtained by this method is consistent with the result of the Bayesian analysis above, within the margin of error. Thus based on current observation of data, there is likely a variability timescale on the order of a few months.

\subsection{Significance Map}

\begin{figure}
\centering
\subfigure[]{\includegraphics[width=0.40\textwidth]{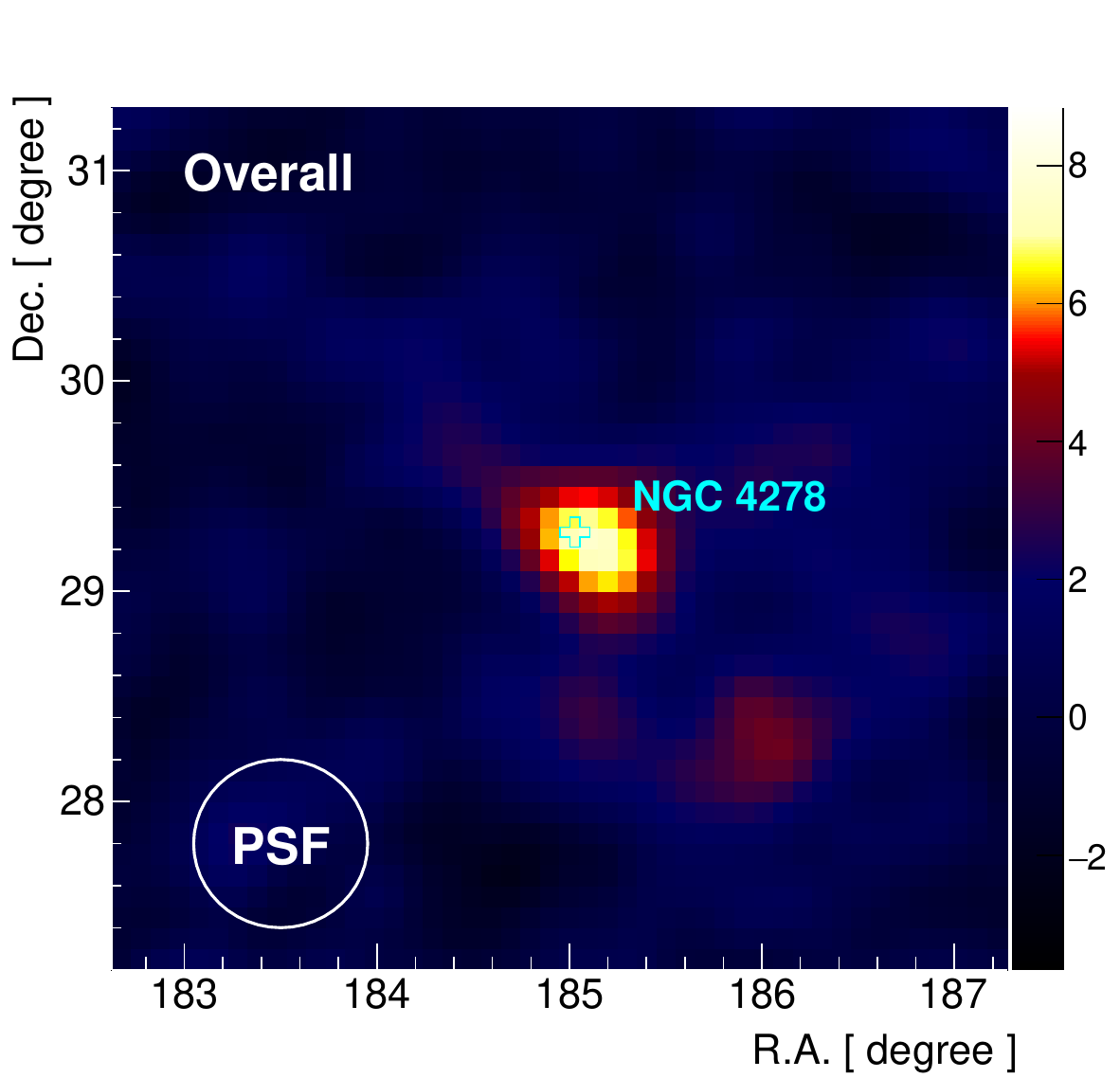}}
\subfigure[]{\includegraphics[width=0.40\textwidth]{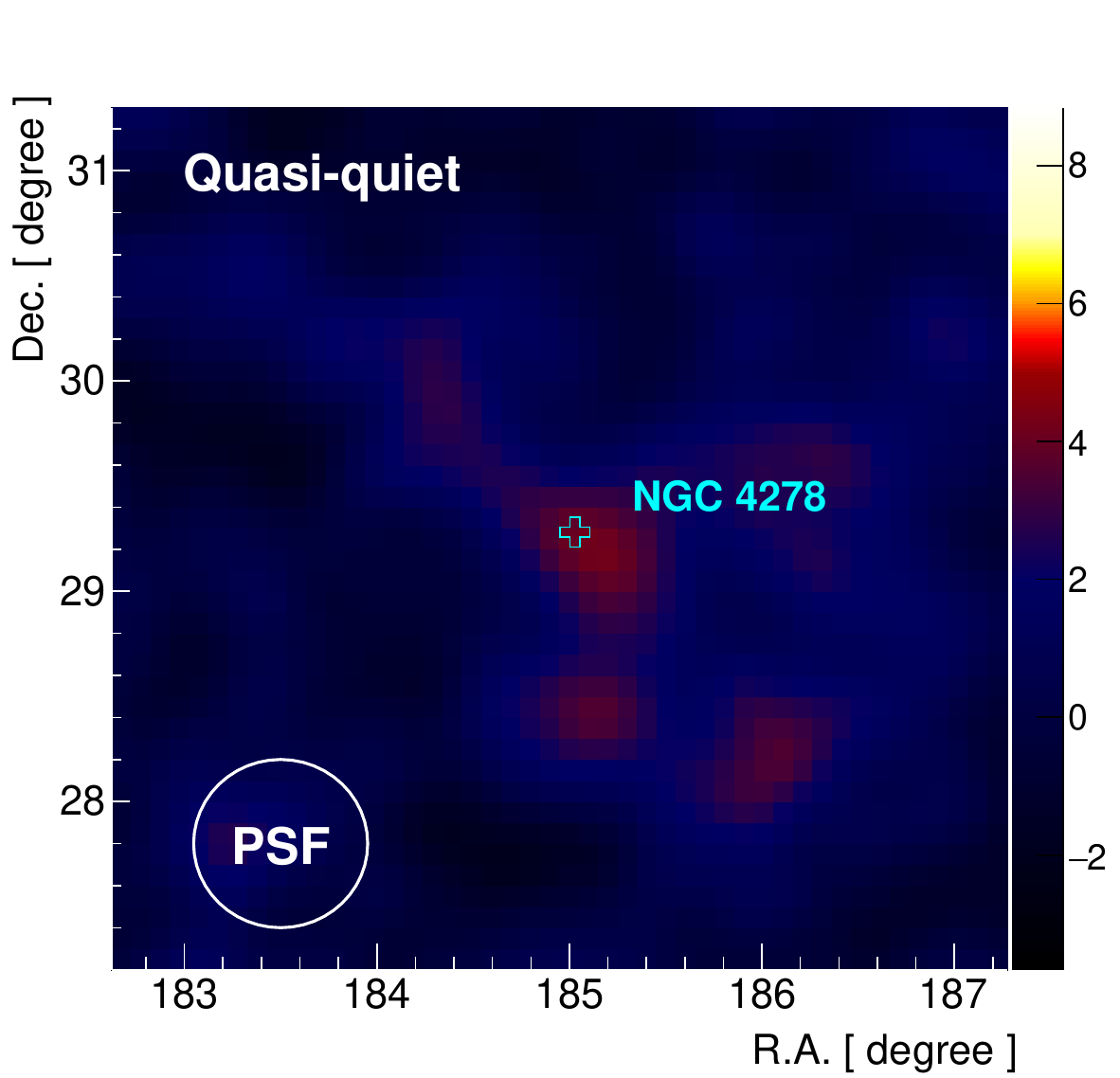}}\\
\subfigure[]{\includegraphics[width=0.40\textwidth]{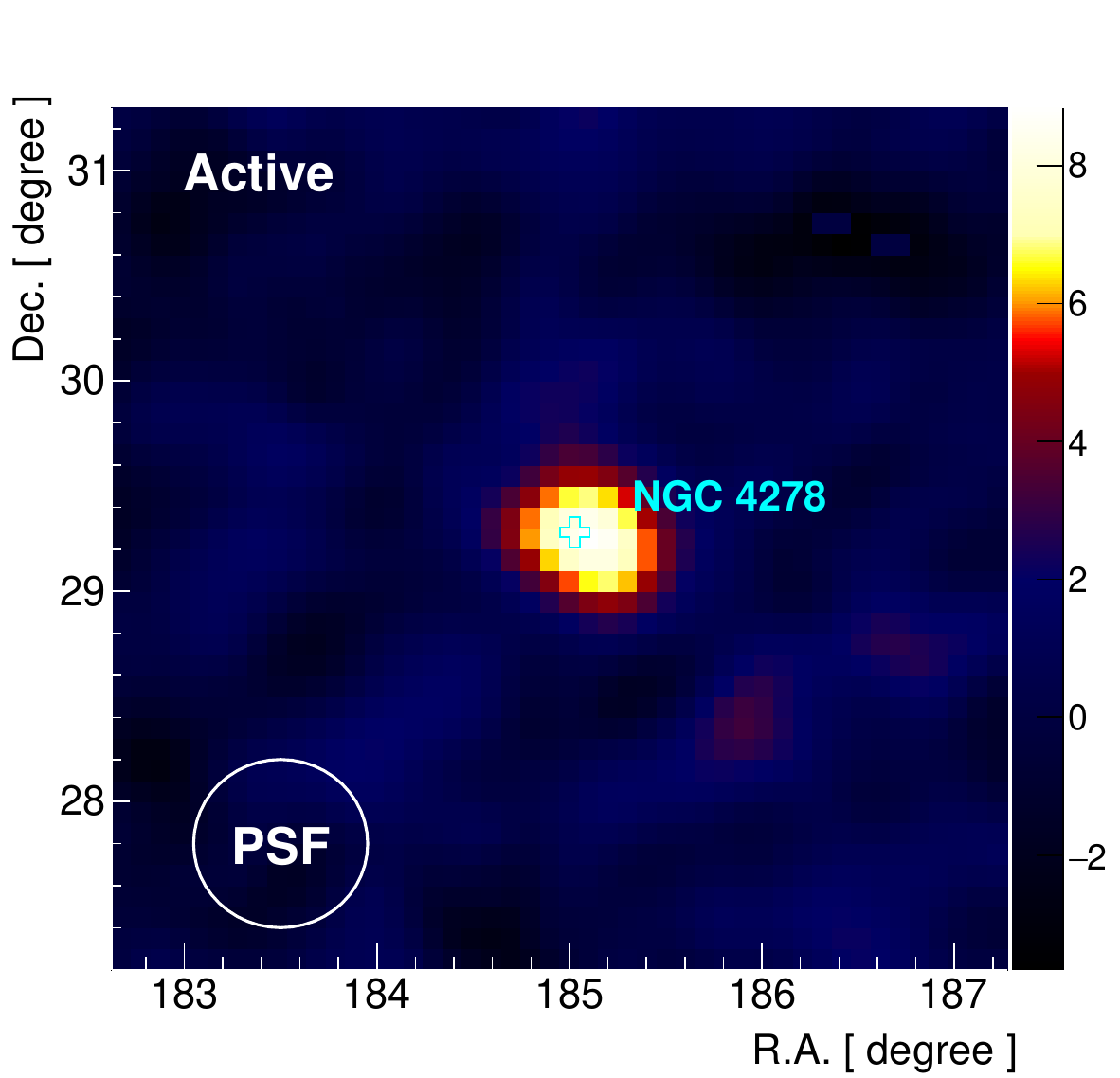}}
\subfigure[]{\includegraphics[width=0.40\textwidth]{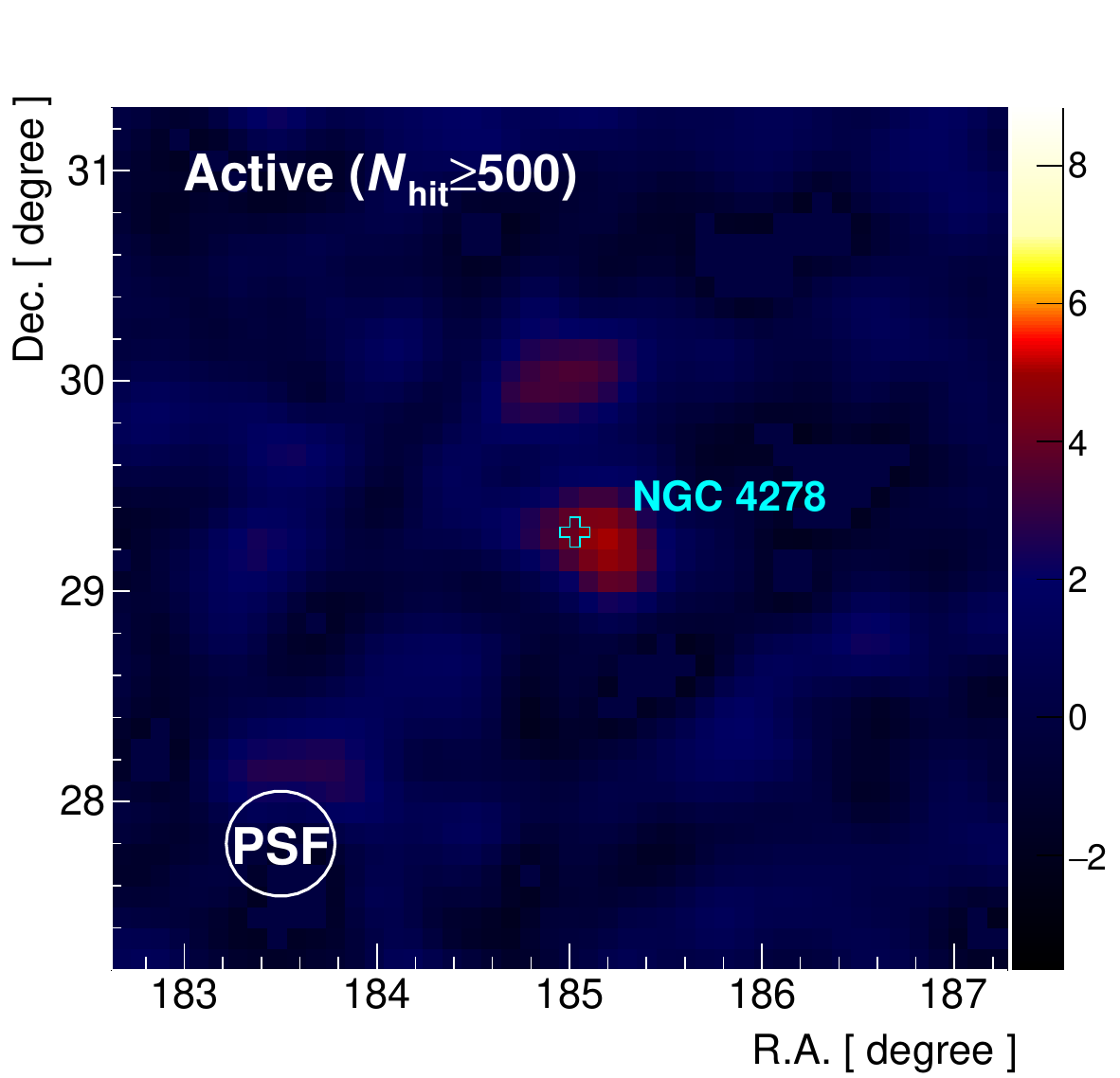}}\\
\caption{The pre-trial significance map around NGC 4278 region by WCDA at different states. Top left is the map with $N_{\rm{hit}}\ge60$ for the entire 891 days of observation, top right is for the quasi-quiet time. Bottom left is the map with $N_{\rm{hit}}\ge60$ during the active state, bottom right is map with $N_{\rm{hit}}\ge500$ during the active state.}
\label{fig_ngc4278_sig}
\end{figure}

The top left panel of the Figure \ref{fig_ngc4278_sig} is the significance map for events with $N_{\rm{hit}}$ larger than 60 in all 891\,days of WCDA observation. If photon spectral index is assumed to be 2.62, the median energy of $N_{\rm{hit}}\ge60$ is around 1~TeV, and the source is detected with a pre-trial statistical significance of 7.6\,$\sigma$. Then the trial numbers is estimated as $\Omega/[2\rm{\pi} (1-\cos(0.5^{\circ}))] = 3.5\times10^{4}$, where $\Omega$ represents the solid angle of the LHAASO sky survey range with declination from $-20^{\circ}$ to $80^{\circ}$, and $0.5^{\circ}$ is the point spread function (PSF) in this energy range. After accounting for the trial numbers, the post-trial significance is 6.1\,$\sigma$. 
And during the quasi-quiet state when the source is not active, the pre-trial significance is around 4.2\,$\sigma$ from WCDA observation.

In contrast, the bottom panels are the map during active period which is defined by Bayesian block analysis. Right panel is 8.8\,$\sigma$ for $N_{\rm{hit}}\ge60$, left is 5.0\,$\sigma$ for $N_{\rm{hit}}\ge500$ that indicates a median energy higher than 10\,TeV. 
The position from likelihood fitting during the active period using WCDA observation is found to be $\rm{R.A.}=185.05^{\circ}\pm0.04^{\circ}$, $\rm{Dec.}=29.25^{\circ}\pm0.03^{\circ}$ which is $0.03^{\circ}$ from the position of NGC 4278 position \citep{Helmboldt:2006wb}. 

The pre-trial significance for all KM2A observations is less than 4\,$\sigma$. Therefore, in the later analysis of spectral energy distribution, only upper limits are estimated from KM2A data.

\subsection{SED in Different Conditions}

\begin{figure}
\centering
\subfigure[]{\includegraphics[width=0.45\textwidth]{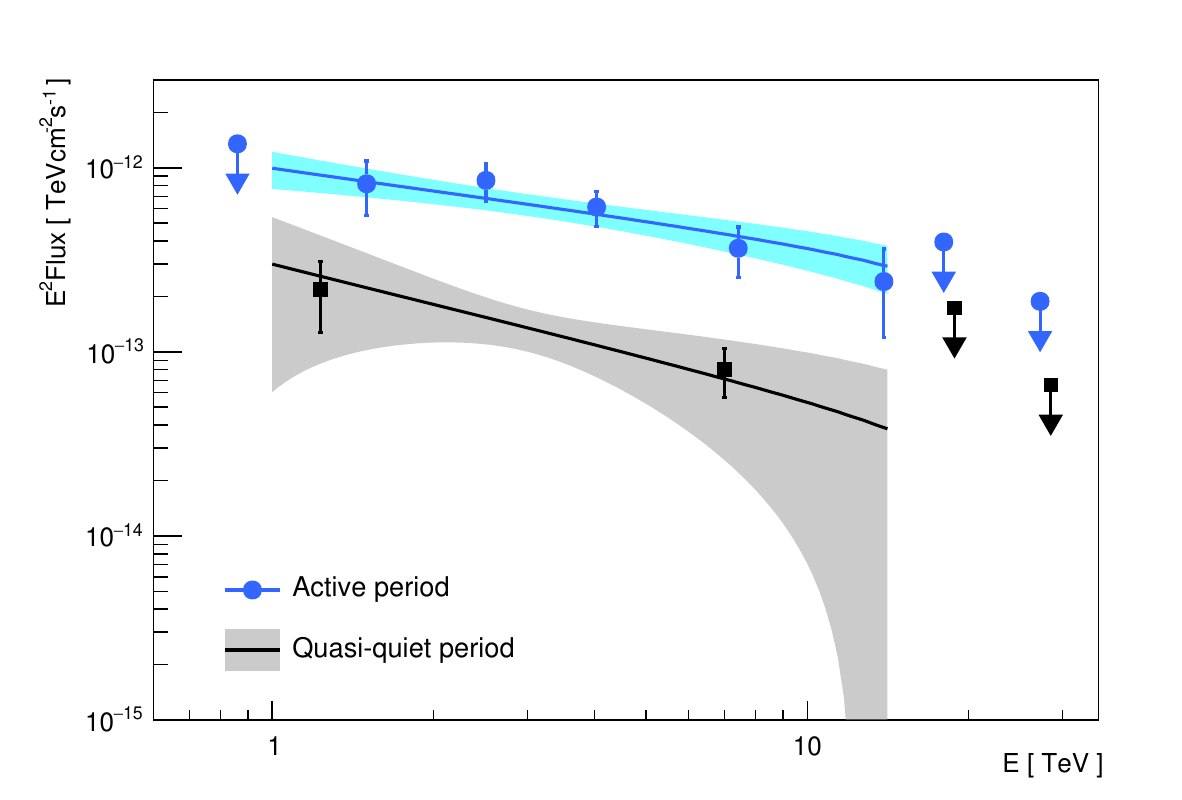}}
\subfigure[]{\includegraphics[width=0.45\textwidth]{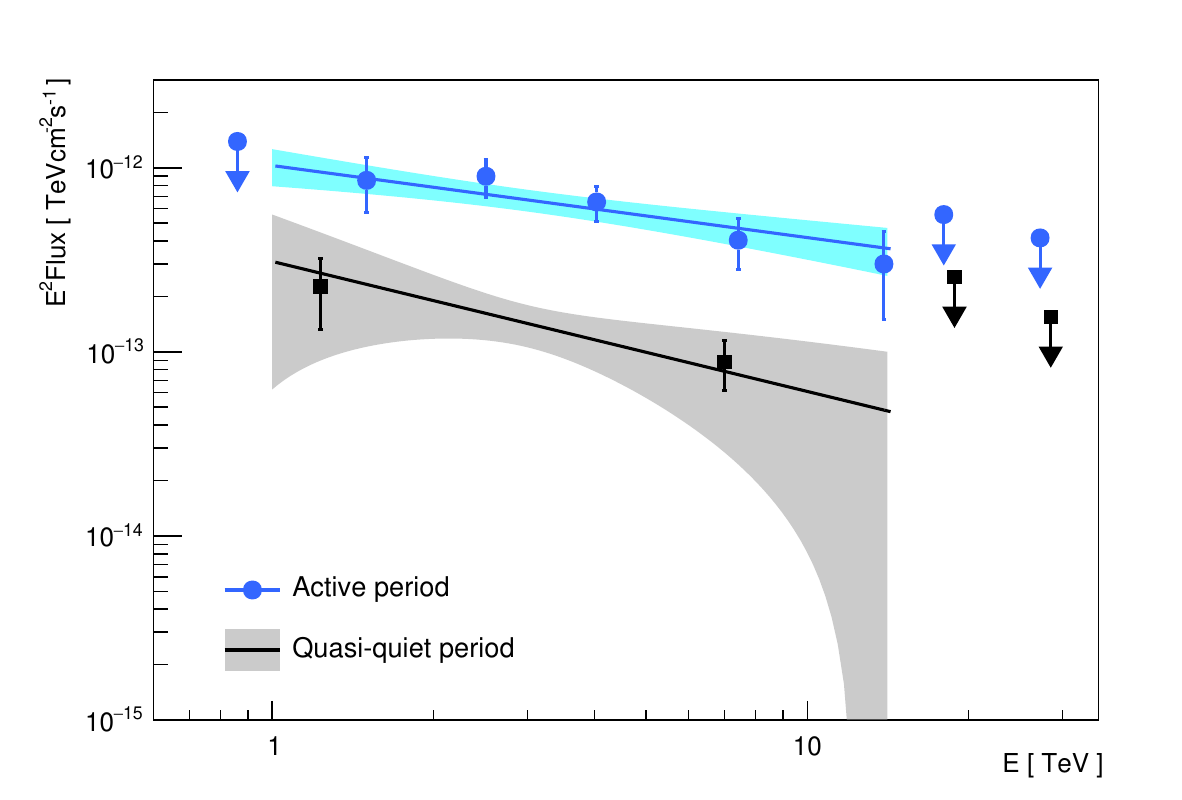}}
\caption{The Observed (left) and intrinsic (right) SED of the source at  different states, blue points is for the SED in the active state, the black line is the SED in the quasi-quiet state. Note that the EBL absorption is not very important due to small distance.}
\label{fig_ngc4278_sed}
\end{figure}

Applying the maximum likelihood method, we obtained the intrinsic spectra for two states: active state which is the period marked by Bayesian block and quasi-quiet state which is the remaining time duration. Figure \ref{fig_ngc4278_sed} shows the observed and intrinsic (corrected for the EBL absorpition) SEDs of the source in these two different states. The SED in the active state is in blue, while that of the quasi-quiet state is shown by black line and grey band. The final two upper limit points are from KM2A measurement. Due to the limited statistics of quasi-quiet state, the 6 segments of data are merged to 2 points as shown in the figure.

It is interesting to see that in both active and quasi-quiet states, there is no obvious change in the intrinsic photon spectral index within the error in the energy range from 1\,TeV to 20\,TeV, the obtained best-fit parameters are $\phi_{0}=(0.74\pm0.10)\times10^{-13}\,\rm{TeV^{-1}cm^{-2}s^{-1}}$, $\varGamma=(2.39\pm0.17)$ with TS value of 78 in active state and $\phi_{0}=(0.16\pm0.04)\times10^{-13}\,\rm{TeV^{-1}cm^{-2}s^{-1}}$, $\varGamma=(2.71\pm0.69$) with TS value of 18.0 in quasi-quiet state.

However, the flux intensity in the active state does increase significantly, at about 7 times higher than that in the quasi-quiet state. Based on the SED measurement of the standard candle, the Crab Nebula, the systematic error in absolute flux is taken as 5\%.

\section{Discussion and Summary}

NGC 4278 exhibits a compact symmetric radio morphology with $\sim3$ pc ``S"-shaped jets \citep{2005ApJ...622..178G}, negligible polarization \citep[$\lesssim0.4\%$,][]{2004MNRAS.352..112B}, months-long variability \citep{2005ApJ...622..178G, 2004MNRAS.352..112B}, low radio luminosity ($\sim10^{38-39}\,\rm{erg\,s^{-1}}$) \citep{2004MNRAS.352..112B, 2005ApJ...622..178G}. VLBA observations reveal a jet proper motion velocity of $v_{\rm{app}}\simeq0.02\text{--}0.2\,\rm{c}$, suggesting a kinetic age $t_{\rm{kin}}\lesssim100\,\rm{years}$ \citep{2005ApJ...622..178G}. These characteristics differ from those of blazars or FRI radio galaxies. As a low-luminosity AGN, NGC 4278 resembles a typical compact symmetric object, making it one of the smallest and youngest radio sources \citep{2005ApJ...622..178G, 2023arXiv230311361R}. Before its detection by LHAASO, all known VHE AGNs were either blazars or FRI radio galaxies with powerful large-scale radio jets. Although some previously known VHE AGNs are LLAGNs (M87 and Cen A), their radio luminosities ($\gtrsim10^{40}\,\rm{erg\,s^{-1}}$) far exceed that of NGC 4278. The observation of VHE TeV emissions from NGC 4278 indicates that even compact, less powerful radio jets can efficiently accelerate particles to emit TeV photons. LHAASO's discovery underscores the importance of survey-mode observations with homogeneous sky coverage in the VHE energy band for identifying TeV sources from various AGN types, especially LLAGNs.

\begin{figure}
\centering
\includegraphics[width=1\textwidth]{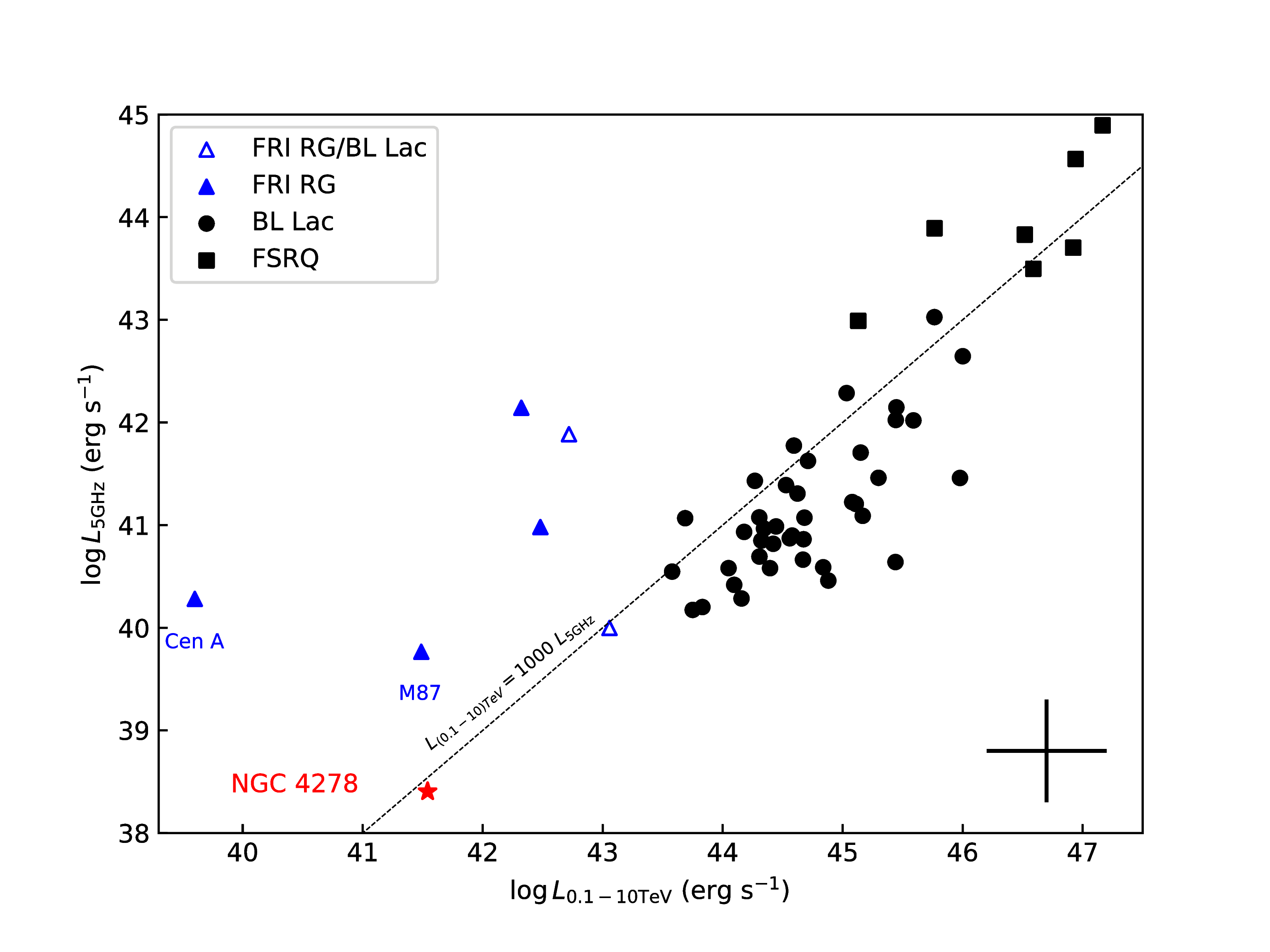}
\caption{Radio luminosity versus TeV luminosity. Data of are obtained from various studies \citep{2014ApJ...797...25P,2009ApJ...695L..40A, 2006Sci...314.1424A,2018MNRAS.476.4187H, 2014A&A...563A..91A, 2022Galax..10...61R,2008Sci...320.1752M, 2013A&A...554A.107H,2011ApJ...730L...8A, 2016A&A...595A..98A,2015ApJ...815L..23A, 2020A&A...633A.162H,2021A&A...647A.163M}. The radio luminosity at 5\,GHz is estimated from observed fluxes and radio spectral indices (assuming $\alpha=0$ for unknown indices).}
\label{TeV AGN}
\end{figure}

Figure \ref{TeV AGN} compares NGC 4278 with other VHE AGNs, plotting radio luminosity against TeV luminosity. The data are from various studies with radio luminosity converted to 5 GHz and VHE luminosity to 0.1-10 TeV. NGC 4278 (red star) has $L_{0.1-10\,\rm{TeV}}\approx 3.0\times10^{41}$ erg s$^{-1}$ and $L_{\rm{5\,GHz}}\approx 2.5\times10^{38}$ erg s$^{-1}$ \citep{2005ApJ...622..178G}. Most VHE AGNs, particularly blazars, exhibit significant variability, can sometime even exceed an order of magnitude. A black cross with $\sigma=0.5$ dex is included in Figure \ref{TeV AGN} to represent luminosity uncertainty. NGC 4278's VHE luminosity is similar to that of FRI RGs despite having a notably lower radio luminosity than that of other VHE AGNs ($L_{\rm{5\,GHz}}\gtrsim10^{40}\,\rm{erg\,s^{-1}}$). NGC 4278 appears to align with the low-luminosity extension of blazars compared to FRI RGs, as guided by the dashed line $L_{0.1-10\,\rm{TeV}}/L_{\rm{5\,GHz}}=1000$. This suggests NGC 4278 efficiently produces TeV emissions akin to blazars, despite its low radio luminosity, slow jet proper motion and less rapid variability. This could be attributed to the possibly small viewing angle of the jet in NGC 4278 \citep{2005ApJ...622..178G}, similar to that in blazars.

\begin{figure}
\centering
\includegraphics[width=1\textwidth]{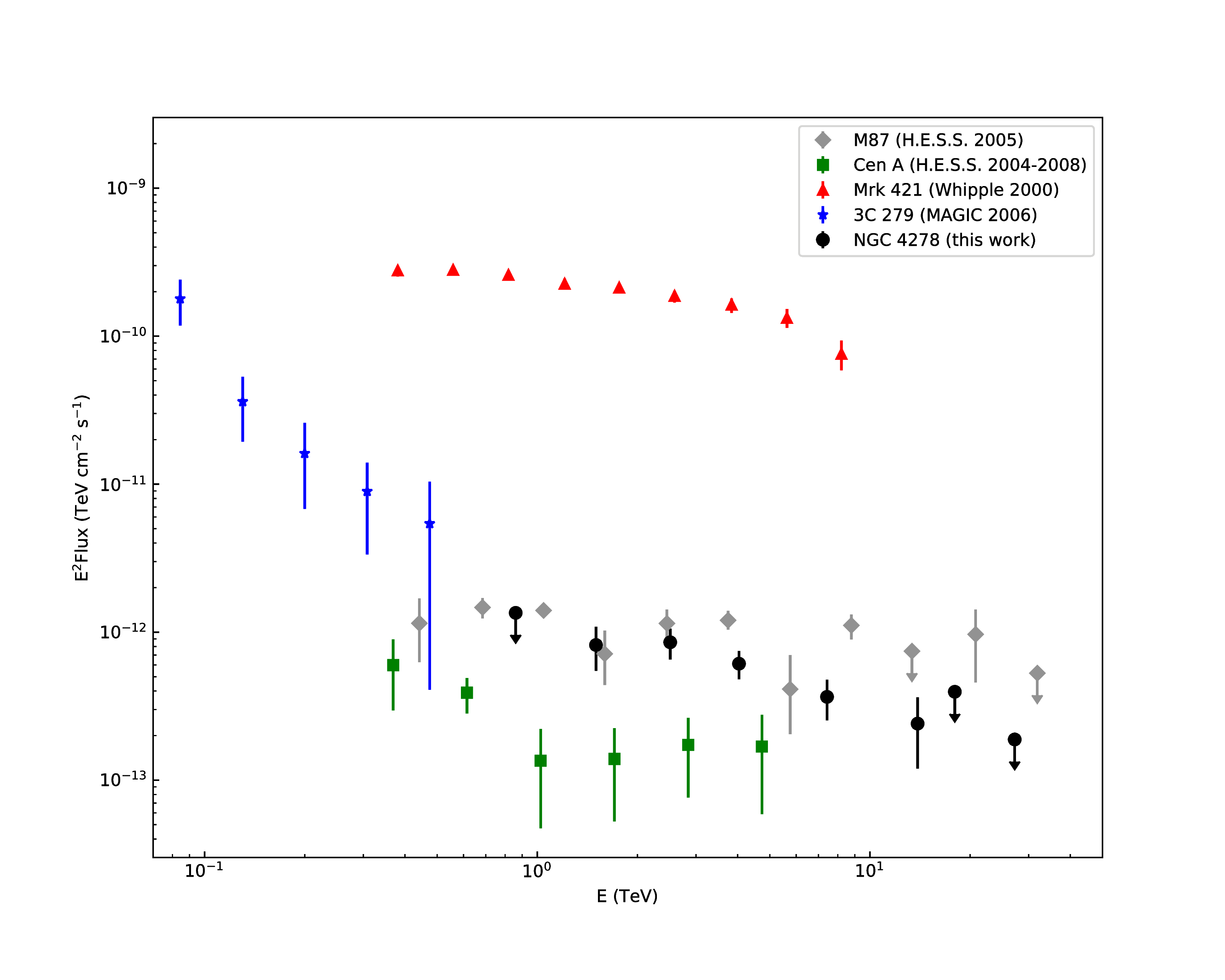}
\caption{The observed VHE spectra of NGC 4278, compare to BL Lac Mrk 421, FSRQ 3C 279, FRI RGs M87 and Cen A, of which
the data are taken during their active periods from \citet{2002ApJ...575L...9K}, \citet{2008Sci...320.1752M}, \citet{2006Sci...314.1424A}, and \citet{2009ApJ...695L..40A}, respectively. It's noteworthy that the extragalactic background light (EBL) significantly absorbs VHE photons above energies of approximately 0.2\,TeV, 9\,TeV, 30\,TeV, 30\,TeV, and 90\,TeV (optical depth $\tau_{\gamma\gamma}=1$) for 3C 279 ($z=0.536$), Mrk 421 ($z=0.03$), M87 ($D_{\rm{l}}\approx16$\,Mpc), NGC 4278 ($D_{\rm{l}}\approx16$\,Mpc), and Cen A ($D_{\rm{l}}=3.8$\,Mpc), respectively \citep[from the EBL model of][]{2021MNRAS.507.5144S}.}
\label{SED VHE AGN}
\end{figure}

The VHE spectra of Mrk 421 (the first discovered VHE BL Lac object), 3C 279 (the first VHE FSRQ), M87 (the first VHE FRI RG), and Cen A (the nearest VHE AGN)  are compared with NGC 4278 in Figure \ref{SED VHE AGN}. VHE emissions from Mrk 421 are typically well interpreted within the synchrotron self-Compton (SSC) model \citep[][]{2015A&A...578A..22A}. While 3C 279's exceed Fermi/LAT GeV extrapolation, possibly from external Compton scattering or hadronic processes \citep{2012ApJ...754..114H, 2009ApJ...703.1168B}. Cen A's VHE emissions may originate from its core and large-scale jet \citep[][]{2020Natur.582..356H}. Although SSC process can explain M87's broadband emissions \citep{2009ApJ...707...55A}, the VHE seems to exceed Fermi/LAT GeV extrapolation, suggesting a structured jet or hadronic origin \citep[][]{2008MNRAS.385L..98T, 2022ApJ...934..158A}. NGC 4278's VHE flux and spectra resemble M87's, with a similar photon index. Fermi/LAT GeV data from the same period as LHAASO observations provide only upper limits, hindering simultaneous broad-band spectra analysis for NGC 4278.

As discussed in Section 3.1, LHAASO observed monthly VHE variability from NGC 4278, with a size ($\lesssim$ light-month) smaller than resolved radio knots \citep{2005ApJ...622..178G}, indicating a possible core origin for its VHE emission. Jet internal shocks within the core can efficiently accelerate particles to emit $\gamma$-rays \citep[e.g.,][]{1978MNRAS.184P..61R}. In the leptonic model, inverse Compton processes responsible for VHE would enter the Klein-Nishina regime, with cooling timescales matching the observed monthly timescale, roughly suggesting $B\gtrsim5\,\rm{mG}$ in the emission region. LLAGNs' collisionless plasma in RIAF could also produce high-energy particles for $\gamma$-ray emission \citep{1997ApJ...490..605M, 2015ApJ...806..159K}, although VHE photons escape from RIAF may be hindered by high opacity \citep{2015ApJ...806..159K}.

Previously known VHE AGNs include blazars and FRI RGs. LHAASO's high sensitivity, wide field of view, and high duty cycle will enhance extragalactic source detection efficiency, reducing biases in VHE AGN catalogs. Data from 891\,days of LHAASO-WCDA observations detected a new TeV source at $\thickapprox8\,\sigma$ \citep[$<4\,\sigma$ from KM2A data, named 1LHAASO J1219+2915 in the first LHAASO catalog,][]{lhaaso-catalog-2024}, which is likely to the AGN in NGC 4278 (within $\sim0.03^{\circ}$). The LHAASO observation unveils moderate variability with monthly variability timescale. The active period flux $f_{1-10\,\rm{TeV}}\approx(7.0\pm1.1_{\rm{sta}} \pm0.35_{\rm{syst}})\times10^{-13}\,\rm{photons}\,\rm{cm}^{-2}\ \rm{s}^{-1}$ ($\sim5\%$ of Crab Nebula, observed photon index $\varGamma=2.56\pm0.14$) is $\sim7$ times higher than its quiet period. The VHE emissions from NGC 4278 suggest that even compact, less powerful radio jets can efficiently accelerate particles to VHE and emit TeV photons. Further LHAASO observations may discover more VHE LLAGNs like NGC 4278, aiding in constructing a comprehensive VHE AGN catalog and understanding AGN physics.

\section{Acknowledgement}
We would like to thank all staff members who work at the LHAASO site above 4400\,m a.s.l. year round to maintain the detector and keep the water recycling system, electricity power supply and other components of the experiment operating smoothly. We are grateful to Chengdu Management Committee of Tianfu New Area for the constant financial support for research with LHAASO data. 
This research work is also supported by the following grants: the National Natural Science Foundation of China (NSFC grants No. 12393853, No. 12393854, No. 12393851, No. 12393852, No. 12173066, No. 12173039), the Department of Science and Technology of Sichuan Province, China (Grant No. 24NSFSC2319) and in Thailand by the National Science and Technology Development Agency (NSTDA) and National Research Council of Thailand (NRCT): High-Potential Research Team Grant Program (N42A650868).

\section{author contribution}

M.~Zha and L.~Chen led the drafting of text and coordinated the whole data analysis, S.C.~Hu performed energy spectrum analysis,  G.M.~Xiang (supervised by M.~Zha) performed light curve analysis,  T.~Wen calculated the upper limits from KM2A, L.~Chen contributed to the interpretation of the data and the studying at multi-wavelength observation, C.D.~Gao provided the cross-check. All other authors participated in data analysis, including detector calibration, data processing, event reconstruction, data quality check, and various simulations, and provided comments on the manuscript.


\clearpage
\bibliographystyle{aasjournal}



\end{document}

%% file: ApJ20240201.tex
\author{Zhen Cao}
\affiliation{Key Laboratory of Particle Astrophysics \& Experimental Physics Division \& Computing Center, Institute of High Energy Physics, Chinese Academy of Sciences, 100049 Beijing, China}
\affiliation{University of Chinese Academy of Sciences, 100049 Beijing, China}
\affiliation{Tianfu Cosmic Ray Research Center, 610000 Chengdu, Sichuan,  China}
 
\author{F. Aharonian}
\affiliation{Dublin Institute for Advanced Studies, 31 Fitzwilliam Place, 2 Dublin, Ireland }
\affiliation{Max-Planck-Institut for Nuclear Physics, P.O. Box 103980, 69029  Heidelberg, Germany}
 
\author{Axikegu}
\affiliation{School of Physical Science and Technology \&  School of Information Science and Technology, Southwest Jiaotong University, 610031 Chengdu, Sichuan, China}
 
\author{Y.X. Bai}
\affiliation{Key Laboratory of Particle Astrophysics \& Experimental Physics Division \& Computing Center, Institute of High Energy Physics, Chinese Academy of Sciences, 100049 Beijing, China}
\affiliation{Tianfu Cosmic Ray Research Center, 610000 Chengdu, Sichuan,  China}
 
\author{Y.W. Bao}
\affiliation{School of Astronomy and Space Science, Nanjing University, 210023 Nanjing, Jiangsu, China}
 
\author{D. Bastieri}
\affiliation{Center for Astrophysics, Guangzhou University, 510006 Guangzhou, Guangdong, China}
 
\author{X.J. Bi}
\affiliation{Key Laboratory of Particle Astrophysics \& Experimental Physics Division \& Computing Center, Institute of High Energy Physics, Chinese Academy of Sciences, 100049 Beijing, China}
\affiliation{University of Chinese Academy of Sciences, 100049 Beijing, China}
\affiliation{Tianfu Cosmic Ray Research Center, 610000 Chengdu, Sichuan,  China}
 
\author{Y.J. Bi}
\affiliation{Key Laboratory of Particle Astrophysics \& Experimental Physics Division \& Computing Center, Institute of High Energy Physics, Chinese Academy of Sciences, 100049 Beijing, China}
\affiliation{Tianfu Cosmic Ray Research Center, 610000 Chengdu, Sichuan,  China}
 
\author{W. Bian}
\affiliation{Tsung-Dao Lee Institute \& School of Physics and Astronomy, Shanghai Jiao Tong University, 200240 Shanghai, China}
 
\author{A.V. Bukevich}
\affiliation{Institute for Nuclear Research of Russian Academy of Sciences, 117312 Moscow, Russia}
 
% ...
\collaboration{288}{The LHAASO Collaboration}